\documentclass[sigconf]{sections/acmart}
\renewcommand\footnotetextcopyrightpermission[1]{}
\setcopyright{none}
\AtBeginDocument{%
  }

\usepackage{amsmath}
\usepackage{mathtools}
\usepackage{enumitem}
\usepackage{booktabs}
\usepackage{makecell}
\usepackage{multirow}
\usepackage{tabularx}
\usepackage{array}
\usepackage{xcolor}
\usepackage{pifont}

\AtEndEnvironment{table}{\vspace{-0.10in}}
\AtEndEnvironment{table*}{\vspace{-0.10in}}

\AtEndPreamble{%
  \hypersetup{
    colorlinks   = true,    
    urlcolor     = blue,    
    linkcolor    = blue,    
    citecolor    = blue     
  }
}

\newenvironment{icompact}{
  \begin{list}{$\bullet$}{
    \itemindent -.05em
    \parsep 0pt plus 1pt
    \partopsep 0pt plus 1pt
    \topsep 2pt plus 2pt minus 2pt
    \itemsep 0pt plus 1.3pt
    \parskip 0pt plus 2pt
    \leftmargin 0.13in}
      }
{\normalsize
\end{list}
}

\begin{document}

\title{WADBERT: Dual-Channel Web Attack Detection via Permutation-Invariant Parameter Modeling}

\author{Kangqiang Luo}
\email{luokangqiang2003@163.com}
\affiliation{%
  \institution{Guangzhou Institute of Technology, Xidian University}
  \city{Guangzhou}
  \country{China}
}

\author{Yi Xie}
\email{yi-xie@tsinghua.edu.cn}
\affiliation{%
  \institution{Tsinghua University}
  \city{Beijing}
  \country{China}
}

\author{Shiqian Zhao}
\email{shiqian.zhao@ntu.edu.sg}
\affiliation{%
  \institution{Nanyang Technological University}
  \city{Singapore}
  \country{Singapore}
}

\author{Jing Pan}
\correspondingauthor
\email{jinglap@aliyun.com}
\affiliation{%
  \institution{Guangzhou Institute of Technology, Xidian University}
  \city{Guangzhou}
  \country{China}
}

\renewcommand{\shortauthors}{Luo et al.}

\begin{abstract}
Web attack detection is the first line of defense for securing web applications, designed to preemptively identify malicious activities. Deep learning-based approaches are increasingly popular for their advantages: automatically learning complex patterns and extracting semantic features from HTTP requests to achieve superior detection performance. However, existing methods are less effective at representing irregular HTTP requests and introduce unnecessary sensitivity to payload parameter order. To tackle this issue, in this paper, we propose WADBERT, a structure-aware dual-channel framework for web attack detection built on two complementary designs. First, we introduce \textit{Hybrid Granularity Embedding (HGE)}, which combines WordPiece embeddings with token-aligned character representations to capture fine-grained patterns in symbol-dense URLs and payload parameters. Second, we develop \textit{structure-aware feature extraction}, with separate branches for the sequential URL and the unordered collection of payload parameters. Each parameter is encoded independently, and position-free multi-head attention followed by mean pooling produces a payload representation invariant to parameter reordering. The URL and payload features are fused for request classification, while the learned parameter-level attention weights support inspection of potentially malicious parameters. Experiments on CSIC2010 and SR-BH2020 validate the effectiveness of our method: WADBERT achieves F1-scores of 99.64\% and 99.59\%, respectively, outperforming the evaluated state-of-the-art baselines.
\end{abstract}





\maketitle
\thispagestyle{empty}  
\pagestyle{empty}      

\section{Introduction}

Web attacks comprise a wide range of malicious behaviors targeting web applications, wherein adversaries exploit software vulnerabilities to hijack accounts, exfiltrate sensitive data, spread malware, or disrupt critical services \cite{wessels2024ssrf,chen2025cross}. As an essential first line of defense, web attack detection inspects incoming traffic to thwart threats before they reach backend services \cite{liu2025bacscan}. At its core, this defense identifies malicious patterns by analyzing the content of HTTP requests across three primary components: the request line, request headers, and the request body.

Recent deep learning approaches formulate web attack detection as a sequence classification task \cite{bokolo2023detection,liu2024transurl,liu2025pmanet}. These methods typically concatenate the URL and payload parameters into a flat text sequence and encode it using language models. However, HTTP requests fundamentally differ from conventional natural language texts in their underlying structure. Specifically, an HTTP request is a semi-structured object consisting of a hierarchical URL path and a collection of key--value payload parameters. These components demand distinct modeling paradigms: while URL path components are inherently sequential and order-sensitive, detecting malicious payloads should primarily depend on the semantics of individual parameters rather than their relative positions in the request. However, existing deep learning methods fail to account for this structural dichotomy, exposing two fundamental challenges:

\textbf{(1) Inadequate HTTP request representation:} Mainstream subword tokenizers (e.g., BPE \cite{sennrich_2016} and WordPiece \cite{devlin2019}) are tailored for natural language corpora and perform poorly on web traffic. In HTTP requests, URLs and payloads frequently feature compound identifiers (e.g., \texttt{getUserName}), character-level encodings (e.g., \texttt{\%27} or \texttt{1\%2B}), and syntax-sensitive tokens where minor character variations drastically alter execution semantics. Consequently, generic subword tokenizers over-segment these strings into fragmented, uninformative tokens, hindering the network from learning robust semantic representations.

\textbf{(2) Order-sensitive parameter modeling:} Malicious patterns can reside in arbitrary payload parameters. Swapping the order of key--value pairs alters their sequence position without changing their actual semantics or security implications, \textit{e.g.}, \texttt{"id=9\&msg=OR 1=1"} versus \texttt{"msg=OR 1=1\&id=9"}. Therefore, a robust detector should produce invariant predictions under parameter permutations. Nevertheless, existing models process payload parameters sequentially, introducing spurious positional dependencies that make detectors vulnerable to simple parameter reordering. This limitation calls for a modeling strategy that captures permutation-invariant combinatorial relationships among parameters rather than superficial sequential orders.

To address these challenges, we propose \textbf{WADBERT}, a structure-aware dual-channel framework that decouples and separately encodes the URL path and payload parameters. Specifically, WADBERT operates across four modular phases: (1) generating token representations via \textit{Hybrid Granularity Embedding (HGE)}, which couples subword semantics with fine-grained character-level features; (2) extracting contextual representations using domain-adapted encoders, URLBERT \cite{li2025} and SecBERT \cite{jackaduma}; (3) aggregating parameter-level features through a position-encoding-free multi-head attention mechanism and mean pooling to derive permutation-invariant payload representations; and (4) concatenating the dual-channel features into a linear classifier for final attack detection. Crucially, beyond holistic request classification, WADBERT leverages its learned parameter-level attention weights to pinpoint suspicious parameters, providing fine-grained diagnostic cues that directly facilitate security triage and incident response.

We evaluate WADBERT on two widely used benchmark datasets: CSIC2010 \cite{CSIC2010} and SR-BH2020 \cite{SRBH2020}. Experimental results demonstrate that WADBERT achieves state-of-the-art F1-scores of 99.64\% and 99.59\%, respectively, outperforming competitive baselines. Comprehensive ablation studies confirm the necessity of HGE, the dual-channel design, and the permutation-invariant aggregation module. Furthermore, inference efficiency evaluations show that WADBERT incurs \textit{comparable latency} with standard Transformer baselines. On the malicious parameter localization task, WADBERT attains an F1-score of 98.60\% across 1,576 matched requests from the WebSpotter benchmark \cite{li2026achieving}, and qualitative attention visualizations verify that the model accurately highlights malicious payloads.

The contributions of this paper are summarized as follows.

\begin{icompact}
    \item We propose WADBERT, a structure-aware dual-channel framework that decouples URL and payload modeling to accurately detect web attacks while also enabling interpretable malicious payload parameter localization.
    \item We introduce Hybrid Granularity Embedding (HGE), which effectively represents symbol-dense strings and token variations commonly found in HTTP requests, consistently outperforming standard WordPiece embeddings.
    \item We design a permutation-invariant aggregation strategy that models combinatorial interactions among payload parameters, conferring robustness against adversarial parameter reordering.
    \item We conduct extensive experiments to evaluate WADBERT. The results demonstrate the superior accuracy and competitive efficiency of our method, as well as its stable performance under parameter-reversal perturbations.
\end{icompact}

\section{Background and Related Work}
\label{sec:related_work}

\subsection{HTTP Request}
\label{sec:http-request}

\begin{figure}[!t]
	\centering
    \includegraphics[width=\linewidth]{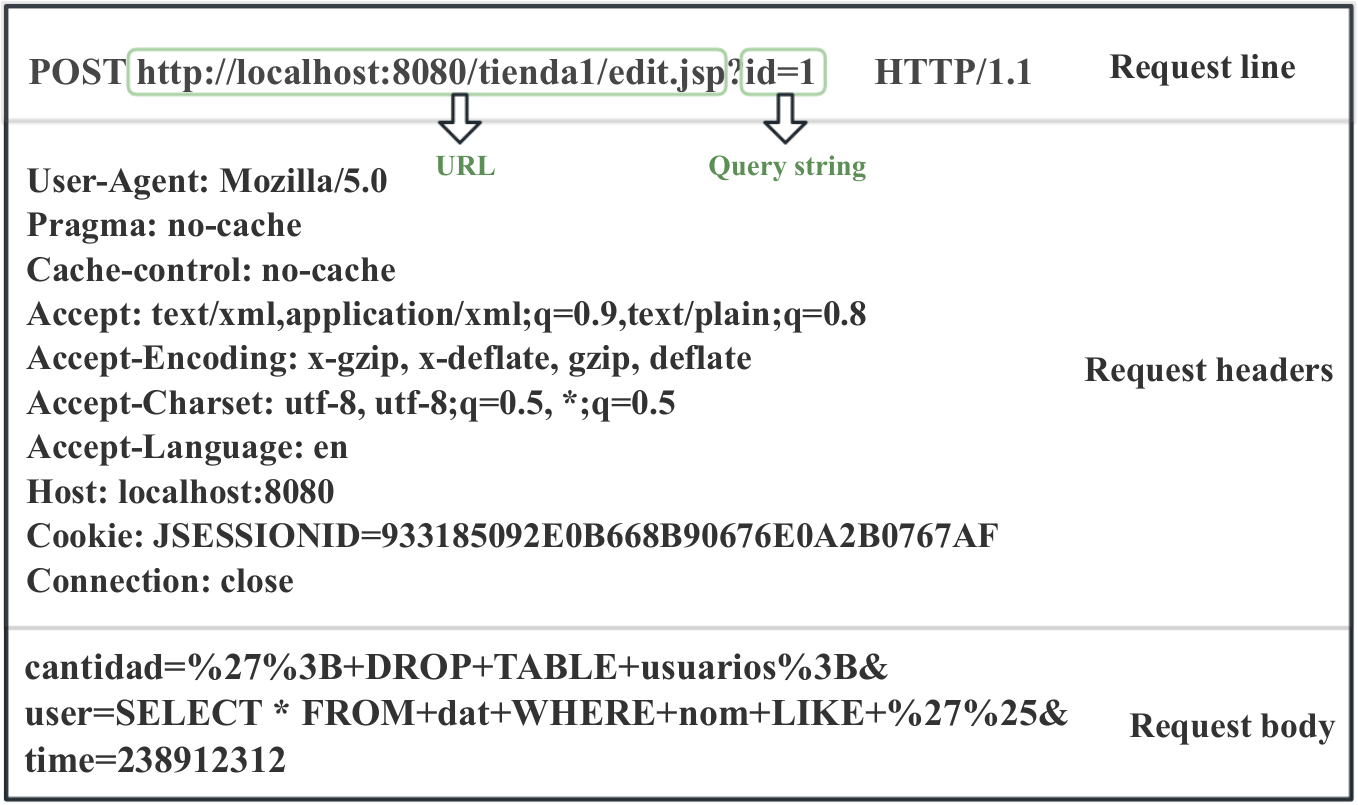}
    \vspace{-0.20in}
	\caption{Components of an HTTP request.}
	\label{fig.HTTP}
    \vspace{-0.10in}
    \Description{}
\end{figure}

HTTP requests enable clients to interact with web servers by requesting resources, submitting data, and invoking application operations. As shown in Figure~\ref{fig.HTTP}, an HTTP request contains an order-sensitive URL and a collection of payload parameters whose ordering should not determine the detection result. We represent a request as $r=(m,u,\mathcal{P})$, where $m$ is the HTTP method, $u$ is the URL path excluding the query string, and $\mathcal{P}=\{\!\{p_1,\ldots,p_n\}\!\}$ is the multiset of $n$ payload parameters extracted from the query string, request headers, and request body. Each $p_k$ is a key--value parameter represented as text. The multiset notation retains parameter multiplicity while making no assumption about parameter order.

\subsection{Web Attacks}
\label{sec:web-attacks}

Web attacks exploit weaknesses in how applications interpret external input and access protected resources. Common classes include SQL injection (SQLi), cross-site scripting (XSS), OS command injection, path traversal, and server-side request forgery (SSRF)~\cite{owasp2020wstg}. SQLi and command injection introduce attacker-controlled instructions into database queries and operating-system commands, respectively. XSS causes attacker-supplied scripts to execute in a victim's browser; the large-scale study by Lekies et al.~\cite{lekies2013flows} demonstrates how unsafe client-side data flows can expose applications to DOM-based XSS. Path traversal manipulates file-related inputs to access resources outside the intended directory. SSRF abuses server-side request functionality to reach attacker-selected destinations. Wessels et al.~\cite{wessels2024ssrf} examine SSRF exploitation and defenses in PHP applications, showing that insecure handling of attacker-controlled request destinations remains a practical concern.

The diversity of input representations further complicates web attack detection. Encoding and syntactic variation can obscure malicious content while retaining its intended behavior. For example, WAF-A-MoLE~\cite{demetrio2020wafamole} uses mutation-based search to construct SQLi variants that evade the evaluated machine learning-based web application firewalls (WAFs). These findings motivate detectors that capture the content and structure of requests beyond isolated surface patterns. WADBERT focuses on attack evidence present in HTTP request content, modeling the URL and individual payload parameters as distinct components.

\begin{figure*}
    \centering
    \includegraphics[width=\linewidth]{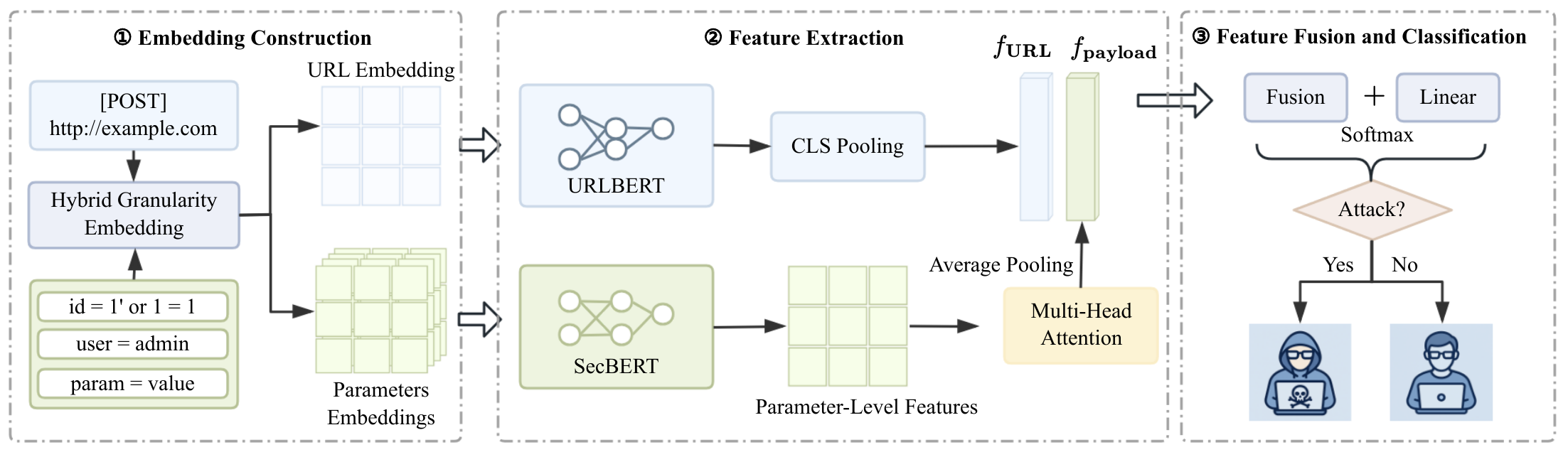}
    \vspace{-0.20in}
    \caption{Overview of WADBERT. HGE constructs hybrid token embeddings for the URL and individual payload parameters. URLBERT and SecBERT encode the two channels, respectively. Position-free parameter-level attention followed by mean pooling produces a permutation-invariant payload feature, which is concatenated with the URL feature for request classification.}
    \label{WADBERT}
    \Description{A dual-channel architecture with hybrid embeddings, separate URL and parameter encoders, parameter-level attention, mean pooling, and a request classifier.}
    \vspace{-0.10in}
\end{figure*}

\subsection{Web Attack Defenses}
\label{sec:web-attack-defenses}

Given the great importance of HTTP requests and the threat of web attacks, it is essential to figure out effective detection methods. In general, within learning-based web attack detection, related work can be grouped into traditional machine learning, traditional deep learning, and Transformer-based methods. 

\noindent \textbf{Machine Learning Methods.}
These methods classify requests using engineered statistical or lexical features~\cite{riera2022new,chakir2023empirical}. Representative classifiers include support vector machines, decision trees, and ensemble models. Chakir et al.~\cite{chakir2023empirical} compare single classifiers with bagging, boosting, and heterogeneous ensembles on web attack datasets, illustrating that detection performance and computational cost depend on the chosen learner and evaluation setting. Such approaches provide practical baselines, but the information available to the classifier is constrained by the selected features. Capturing fine-grained relationships within symbol-dense payloads therefore requires appropriate feature design.

\noindent \textbf{Traditional Deep Learning Methods.}
These approaches learn representations from character or word sequences. Attack-specific studies include DeepXSS~\cite{Fang_2018} for XSS detection and neural SQLi detection~\cite{Tang_2020}. Other work applies convolutional models to web attack detection~\cite{Tekerek_2021} and develops distributed deep learning systems for edge deployment~\cite{Tian_2020}. Ensemble designs such as EDL~\cite{Luo_2021}, together with CNN-BiLSTM models studied in E-WebGuard~\cite{zhou2025webguard}, combine complementary neural representations. These methods reduce reliance on manually specified features. However, encoding a serialized request as one sequence can introduce sensitivity to the arrangement of parameter fields, motivating representations that explicitly account for HTTP structure.

\noindent \textbf{Transformer-Based Methods.}
Contextual language representations have been applied to both web request classification and malicious URL detection. Bokolo et al.~\cite{bokolo2023detection} investigate DistilBERT-based web attack detection, while LogBERT-BiLSTM~\cite{ramos2022logbert} combines BERT representations with bidirectional LSTMs to identify malicious HTTP requests. In malicious URL detection, TransURL~\cite{liu2024transurl} integrates multi-layer Transformer encoding with multi-scale pyramid features, and PMANet~\cite{liu2025pmanet} uses post-training and multi-level feature attention. These URL-focused methods improve contextual feature extraction, while modeling an entire HTTP request additionally requires accounting for the organization of its payload parameters. A sequence encoder alone does not impose invariance to parameter reordering. Locate-Then-Detect~\cite{liu2019locate} first locates suspicious regions and then classifies them. WebSpotter~\cite{li2026achieving} uses field-importance information to localize malicious payloads. Such work provides useful support for attack analysis.

WADBERT focuses on the structure of HTTP requests. It combines character-aware token embeddings with separate URL and payload parameter encoders. Position-free attention and mean pooling jointly yield a payload representation that is invariant to parameter order. The learned attention weights also support the inspection of potentially malicious parameters.

\vspace{-0.09in}
\section{WADBERT}
\label{sec:methodology}
\subsection{Overview}

\begin{figure*}
    \centering
    \includegraphics[width=\linewidth]{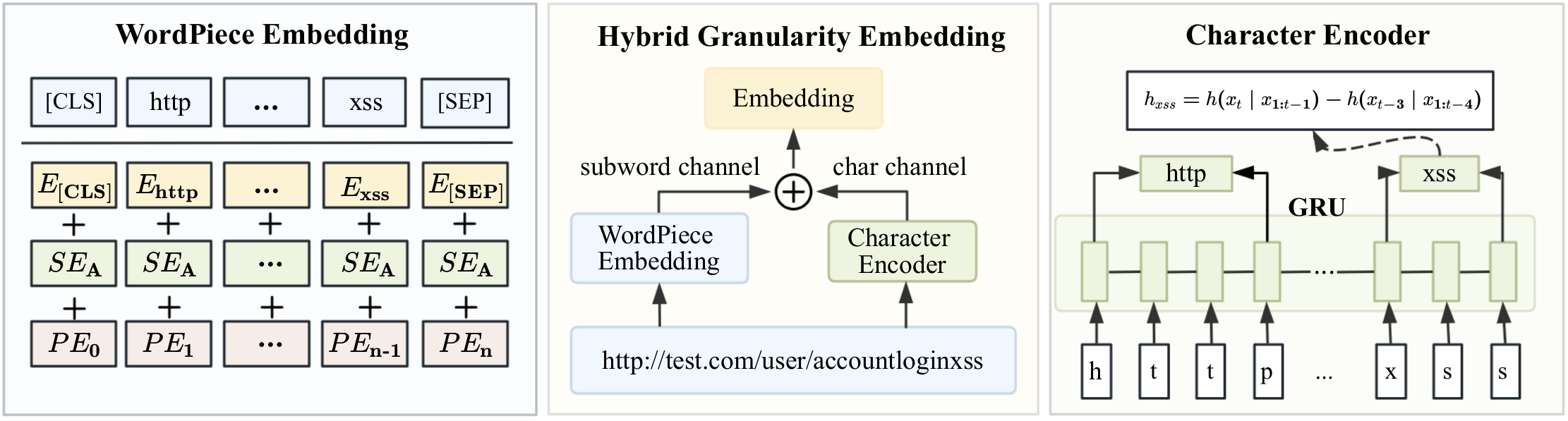}
    \vspace{-0.20in}
    \caption{Framework of HGE. A bidirectional GRU encodes the character sequence. Differential states at each WordPiece token's character boundaries are concatenated and projected into the WordPiece embedding space, then added to the token, position, and segment embeddings.}
    \label{HGE}
    \Description{Character embedding and bidirectional GRU encoding followed by token-span differential representations, projection, and fusion with WordPiece embeddings.}
    \vspace{-0.10in}
\end{figure*}
\noindent \textbf{Motivation.}
Detecting web attacks requires capturing \textit{fine-grained textual cues} while respecting \textit{HTTP request structure}. URLs and payload parameters often contain symbols, encodings, and non-standard strings whose character patterns may be obscured by subword tokenization; character-only tokenization, however, lengthens the encoder input. Meanwhile, URL paths and parameter contents are order-sensitive sequences, whereas our multiset representation (Section~\ref{sec:http-request}) calls for invariance to parameter reordering. Flattening a request into one sequence can conflate these distinct ordering properties. These considerations motivate character-aware embeddings at subword granularity and a structure-aware encoder that preserves local sequence order while aggregating parameters without positional dependence.

\noindent \textbf{Our Design.}
To address these requirements, WADBERT combines \textit{hybrid-granularity embeddings} with \textit{dual-channel feature extraction}: it encodes the URL as a sequence and the payload as a collection of independently encoded parameters. Given a request $r$, it predicts a probability vector $\hat{\mathbf{p}}=(\hat{p}_0,\hat{p}_1)^\top$, where classes $0$ and $1$ denote benign and malicious requests, respectively. As illustrated in Figure~\ref{WADBERT}, the architecture consists of three modules:
\begin{icompact}
    \item \textbf{Hybrid Granularity Embedding.} HGE combines WordPiece embeddings with token-aligned character representations for the URL and each payload parameter.
    \item \textbf{Structure-Aware Feature Extraction.} URLBERT encodes the method and URL path, while SecBERT encodes each parameter independently. Position-free attention captures interactions across these parameters, and subsequent mean pooling yields a parameter-order-invariant payload feature.
    \item \textbf{Feature Fusion and Classification.} A linear classifier followed by a softmax layer predicts the request class from the concatenated URL and payload features.
\end{icompact}

We introduce the detailed design of our method as follows.

\subsection{Hybrid Granularity Embedding (HGE)}
\label{sec:hge}

URLs and payload parameters often contain symbol-dense strings and non-standard tokens. HGE supplements WordPiece representations with character-level information without treating every character as a separate Transformer input token. As illustrated in Figure~\ref{HGE}, it consists of character encoding, token-aligned representation construction, and embedding fusion. We use bold lowercase letters for vectors, bold uppercase letters for matrices, and upright subscripts or superscripts for semantic labels.

\noindent \textbf{Step I: Tokenization and Character Encoding.}
For an input text $x$, let $T=(t_1,\ldots,t_M)$ denote its WordPiece content tokens and $C=(c_1,\ldots,c_L)$ its character sequence. Here, $M$ excludes the special tokens, and $L$ is the character count. A character embedding matrix $\mathbf{W}_{\mathrm{emb}}\in\mathbb{R}^{v_{\mathrm{c}}\times d_{\mathrm{c}}}$ maps each character to
$\mathbf{x}_j\coloneqq\mathbf{W}_{\mathrm{emb}}[\operatorname{id}(c_j),:]^\top\in\mathbb{R}^{d_{\mathrm{c}}}$,
where $v_{\mathrm{c}}$ is the character vocabulary size, $d_{\mathrm{c}}$ is the embedding dimension, and $\operatorname{id}$ maps a character to its vocabulary index. A bidirectional GRU~\cite{cho2014learning} encodes the character sequence:
\begin{align}
\overrightarrow{\mathbf{h}}_j
&=\operatorname{GRU}_{\mathrm{fwd}}
(\mathbf{x}_j,\overrightarrow{\mathbf{h}}_{j-1}),
\label{eq:forward}\\
\overleftarrow{\mathbf{h}}_j
&=\operatorname{GRU}_{\mathrm{bwd}}
(\mathbf{x}_j,\overleftarrow{\mathbf{h}}_{j+1}),
\label{eq:backward}
\end{align}
for $j=1,\ldots,L$. Both hidden states belong to $\mathbb{R}^{d_{\mathrm{h}}}$, where $d_{\mathrm{h}}$ is the hidden dimension per direction.

\noindent \textbf{Step II: Token-Aligned Character Representation.}
For content token $t_i$, let $s_i$ and $e_i$ be the indices of its first and last characters, with $1\leq s_i\leq e_i\leq L$. We use one-based character indices and boundary states
$\overrightarrow{\mathbf{h}}_0=\mathbf{0}_{d_{\mathrm{h}}}$ and
$\overleftarrow{\mathbf{h}}_{L+1}=\mathbf{0}_{d_{\mathrm{h}}}$,
where $\mathbf{0}_{d_{\mathrm{h}}}$ is the zero vector in $\mathbb{R}^{d_{\mathrm{h}}}$. The forward and backward differential representations are
\begin{align}
\mathbf{h}_i^{\mathrm{fw}}
&\coloneqq\overrightarrow{\mathbf{h}}_{e_i}
-\overrightarrow{\mathbf{h}}_{s_i-1},
\label{eq:fwd-diff}\\
\mathbf{h}_i^{\mathrm{bw}}
&\coloneqq\overleftarrow{\mathbf{h}}_{s_i}
-\overleftarrow{\mathbf{h}}_{e_i+1}.
\label{eq:bwd-diff}
\end{align}
These expressions also cover tokens at either boundary. Concatenating the two directions gives
\begin{equation}
\mathbf{h}_i^{\mathrm{char}}
\coloneqq
\begin{bmatrix}
\mathbf{h}_i^{\mathrm{fw}}\\
\mathbf{h}_i^{\mathrm{bw}}
\end{bmatrix}
\in\mathbb{R}^{2d_{\mathrm{h}}}.
\label{eq:char-concatenation}
\end{equation}
Thus, each WordPiece content token receives a character-level representation aligned with its span in the input text.

\noindent \textbf{Step III: Hybrid Embedding Fusion.}
A learned projection maps each character representation into the encoder's $d$-dimensional embedding space:
\begin{equation}
\mathbf{e}_i^{\mathrm{char}}
\coloneqq\operatorname{Proj}_{\mathrm{char}}
(\mathbf{h}_i^{\mathrm{char}})
\in\mathbb{R}^{d},
\label{eq:char-projection}
\end{equation}
where $\operatorname{Proj}_{\mathrm{char}}:\mathbb{R}^{2d_{\mathrm{h}}}\rightarrow\mathbb{R}^{d}$. At the same content-token position, the WordPiece embedding is
$\mathbf{e}_i^{\mathrm{WP}}
\coloneqq\mathbf{e}_i^{\mathrm{tok}}+\mathbf{e}_i^{\mathrm{pos}}+\mathbf{e}_i^{\mathrm{seg}}$,
the sum of the token, position, and segment embeddings. Let
$\mathbf{E}_{\mathrm{WP}},\mathbf{E}_{\mathrm{char}}\in\mathbb{R}^{M\times d}$
stack these WordPiece and character embeddings as rows, respectively. The fused content-token embeddings are
\begin{equation}
\mathbf{E}
\coloneqq\mathbf{E}_{\mathrm{WP}}
+\mathbf{E}_{\mathrm{char}}
\in\mathbb{R}^{M\times d}.
\label{eq:sum-char-wp}
\end{equation}

The equations above specify fusion at content-token positions. We use $\operatorname{HGE}(x)$ to denote the complete encoder input construction, including \texttt{[CLS]} and \texttt{[SEP]} at the beginning and end. If $x$ has $M_x$ content tokens, this input has $M_x+2$ rows. WADBERT uses the same embedding construction in both channels, with embeddings compatible with the corresponding encoder. Token-level positional embeddings are retained within each text; positional encoding is omitted only when aggregating across payload parameters.

\subsection{Structure-Aware Feature Extraction}
\label{sec:structure-aware-feature-extraction}
\label{sec:URL-feature-extraction}
\label{sec:payload-feature-extraction}

\noindent \textbf{URL Feature Extraction.}
The URL branch preserves sequential information in the URL path while incorporating the HTTP method. It consists of preprocessing, HGE embedding construction, and contextual encoding with URLBERT~\cite{li2025}.

\noindent \textit{Data Preprocessing.}
Let $\bar{u}\coloneqq\operatorname{Prep}_{\mathrm{URL}}(m,u)$ denote the URL-branch input. The preprocessing operator collapses consecutive slashes in the path into a single slash, converts the path to lowercase, and prepends the HTTP method (e.g., \texttt{GET} or \texttt{POST}). Including the method allows the encoder to distinguish requests that access the same path through different operations.

\noindent \textit{Embedding and Contextual Encoding.}
Let $M_u$ be the number of WordPiece content tokens in $\bar{u}$. HGE first constructs the input embeddings, including \texttt{[CLS]} and \texttt{[SEP]}, and URLBERT then produces the contextual states:
\begin{equation}
\begin{aligned}
\mathbf{E}_{\mathrm{URL}}
&\coloneqq\operatorname{HGE}(\bar{u}),\\
\mathbf{H}_{\mathrm{URL}}
&=\operatorname{URLBERT}(\mathbf{E}_{\mathrm{URL}}),
\end{aligned}
\label{eq:url-encoding}
\end{equation}
where both matrices belong to $\mathbb{R}^{(M_u+2)\times d}$. We represent feature vectors as columns and token states as matrix rows. The global URL feature is therefore
$\mathbf{f}_{\mathrm{URL}}\coloneqq
(\mathbf{H}_{\mathrm{URL}})_{1,:}^{\top}\in\mathbb{R}^{d}$,
the final-layer state of the first token, \texttt{[CLS]}.

\noindent \textbf{Payload Feature Extraction.}
The payload branch retains token order within each parameter while aggregating information across different parameters without using their positions in the HTTP request. It consists of independent preprocessing and encoding, followed by parameter-level fusion.

\noindent \textit{Data Preprocessing.}
For each $p_k$, we apply recursive URL decoding until the decoded text no longer changes, followed by Unicode normalization that converts full-width characters to half-width characters. For example, recursive decoding maps \texttt{\%253C} to \texttt{<}. Invalid or incomplete encodings are preserved. We denote the result by
$\bar{p}_k\coloneqq\operatorname{Prep}_{\mathrm{P}}(p_k)$.
This operation is applied independently to each parameter, so changing their enumeration does not change their individual preprocessed texts.

\noindent \textit{Per-Parameter Encoding.}
Let $M_k$ denote the number of WordPiece content tokens in $\bar{p}_k$. SecBERT~\cite{jackaduma} independently encodes each parameter's hybrid embeddings:
\begin{equation}
\begin{aligned}
\mathbf{E}_k^{\mathrm{P}}
&\coloneqq\operatorname{HGE}(\bar{p}_k),\\
\mathbf{H}_k^{\mathrm{P}}
&=\operatorname{SecBERT}(\mathbf{E}_k^{\mathrm{P}}),\\
\mathbf{z}_k
&\coloneqq(\mathbf{H}_k^{\mathrm{P}})_{1,:}^{\top},
\end{aligned}
\label{eq:parameter-encoding}
\end{equation}
where $\mathbf{E}_k^{\mathrm{P}},\mathbf{H}_k^{\mathrm{P}}
\in\mathbb{R}^{(M_k+2)\times d}$ and
$\mathbf{z}_k\in\mathbb{R}^{d}$ is the final-layer \texttt{[CLS]} feature. The same SecBERT encoder is used for all parameters. Stacking their features as rows gives
\begin{equation}
\mathbf{P}
\coloneqq[\mathbf{z}_1,\ldots,\mathbf{z}_n]^\top
\in\mathbb{R}^{n\times d}.
\label{eq:parameter-matrix}
\end{equation}
The row indices enumerate parameters for computation; they are not supplied as positional features.

\noindent \textit{Parameter-Level Fusion.}
A multi-head self-attention module captures interactions among the parameter features without adding parameter-level positional encodings:
\begin{equation}
\mathbf{Z}=\operatorname{MHA}(\mathbf{P})
=[\mathbf{a}_1,\ldots,\mathbf{a}_n]^\top
\in\mathbb{R}^{n\times d},
\label{eq:parameter-attention}
\end{equation}
where $\mathbf{a}_k\in\mathbb{R}^{d}$ is the attention-enhanced representation of parameter $k$. For $n\geq1$, define the aggregation map $\rho:\mathbb{R}^{n\times d}\rightarrow\mathbb{R}^{d}$ by mean pooling:
\begin{equation}
\mathbf{f}_{\mathrm{payload}}\coloneqq\rho(\mathbf{P})
=\frac{1}{n}\mathbf{Z}^\top\mathbf{1}_n
=\frac{1}{n}\sum_{k=1}^{n}\mathbf{a}_k,
\label{eq:mean-pooling}
\end{equation}
where $\mathbf{1}_n$ is the all-ones vector in $\mathbb{R}^{n}$.

For any permutation matrix $\mathbf{R}\in\{0,1\}^{n\times n}$, position-free self-attention satisfies
$\operatorname{MHA}(\mathbf{R}\mathbf{P})
=\mathbf{R}\operatorname{MHA}(\mathbf{P})$.
Reordering the parameters therefore reorders the attention outputs, while their mean remains unchanged:
\begin{equation}
\rho(\mathbf{R}\mathbf{P})=\rho(\mathbf{P}).
\label{eq:payload-invariance}
\end{equation}
Appendix~\ref{app:permutation_invariance} provides the proof. This property concerns reordering a fixed collection of retained parameters under deterministic inference. It does not remove token-order information within a parameter or within the URL.

\noindent \textit{Attention-Based Parameter Localization.}
The attention weights also provide diagnostic scores for inspecting individual parameters. Let $\mathbf{A}^{(h)}\in[0,1]^{n\times n}$ be the row-normalized attention matrix of head $h$, where $A_{\ell k}^{(h)}$ is the weight from query parameter $\ell$ to parameter $k$. With $H$ attention heads, define the received-attention score
\begin{equation}
s_k\coloneqq\frac{1}{Hn}
\sum_{h=1}^{H}\sum_{\ell=1}^{n}A_{\ell k}^{(h)}.
\label{eq:parameter-importance}
\end{equation}
Because $\sum_{k=1}^{n}s_k=1$, the request-adaptive threshold $1/n$ identifies parameters with $s_k\geq1/n$ as potentially malicious. This rule supports multiple flagged parameters and requires no additional localization model. These scores are diagnostic cues rather than causal explanations of the classifier's decisions; the localization evaluation is reported in Section~\ref{sec:evaluation}.

\subsection{Feature Fusion and Classification}
\label{sec:classification}

The two branches provide complementary request-level features. Their column-vector concatenation is
\begin{equation}
\mathbf{f}\coloneqq
\begin{bmatrix}
\mathbf{f}_{\mathrm{URL}}\\
\mathbf{f}_{\mathrm{payload}}
\end{bmatrix}
\in\mathbb{R}^{2d}.
\label{Eq.6}
\end{equation}
A linear classifier followed by softmax predicts the probabilities of the benign and malicious classes:
\begin{equation}
\hat{\mathbf{p}}
=\operatorname{softmax}
(\mathbf{W}_{\mathrm{cls}}\mathbf{f}
+\mathbf{b}_{\mathrm{cls}}),
\label{Eq.7}
\end{equation}
where $\mathbf{W}_{\mathrm{cls}}\in\mathbb{R}^{2\times2d}$ and
$\mathbf{b}_{\mathrm{cls}}\in\mathbb{R}^{2}$ are trainable parameters. The probabilities satisfy
$\hat{p}_0+\hat{p}_1=1$, and the predicted class is
$\hat{y}\coloneqq\arg\max_{c\in\{0,1\}}\hat{p}_c$.

For a ground-truth class $y\in\{0,1\}$, let
$\mathbf{y}=(y_0,y_1)^\top\in\{0,1\}^{2}$ be its one-hot encoding: $y_c=1$ if $c=y$ and $0$ otherwise. The per-request cross-entropy loss is
\begin{equation}
\mathcal{L}(r,y)\coloneqq
-\sum_{c=0}^{1}y_c\log\hat{p}_c
=-\log\hat{p}_y.
\label{Eq.8}
\end{equation}
Training minimizes this loss over the labeled requests. Reordering payload parameters preserves both branch features. The resulting probability vector $\hat{\mathbf{p}}$ is therefore unchanged.

\section{Experiments}
\label{sec:evaluation}

\subsection{Experiment Setup}
\label{subsec:exp_settings}
\begin{table*}[t]
\centering
\caption{Performance comparison of different methods
on original and parameter-reversed test sets (\%).
$\Delta$F1 denotes the decrease in F1-score in percentage points,
while RI denotes the relative F1-score improvement of
WADBERT over each baseline on the reversed test set (\%).}
\label{tab:overall-performance}
\setlength{\tabcolsep}{2pt}
\renewcommand{\arraystretch}{1.1}
\resizebox{\textwidth}{!}{%
\begin{tabular}{l*{20}{c}}
\toprule
& \multicolumn{10}{c}{CSIC2010}
& \multicolumn{10}{c}{SR-BH2020} \\
\cmidrule(lr){2-11}
\cmidrule(lr){12-21}

& \multicolumn{4}{c}{Original}
& \multicolumn{4}{c}{Reversed}
& \multicolumn{2}{c}{Comparison}
& \multicolumn{4}{c}{Original}
& \multicolumn{4}{c}{Reversed}
& \multicolumn{2}{c}{Comparison} \\
\cmidrule(lr){2-5}
\cmidrule(lr){6-9}
\cmidrule(lr){10-11}
\cmidrule(lr){12-15}
\cmidrule(lr){16-19}
\cmidrule(lr){20-21}

Method
& Acc. & Pre. & Rec. & F1
& Acc. & Pre. & Rec. & F1
& $\Delta$F1 & RI
& Acc. & Pre. & Rec. & F1
& Acc. & Pre. & Rec. & F1
& $\Delta$F1 & RI \\
\midrule

CNN-BiLSTM
& 96.72 & 96.19 & 95.81 & 96.00
& 88.01 & 81.74 & 91.14 & 86.19
& 9.81 & 15.61
& 96.27 & 98.22 & 96.27 & 97.24
& 84.33 & 85.25 & 93.14 & 89.02
& 8.22 & 11.87 \\

EDL
& 99.13 & 99.45 & 98.42 & 98.93
& 88.45 & 78.58 & 98.79 & 87.53
& 11.40 & 13.84
& 96.67 & 99.48 & 95.62 & 97.51
& 88.53 & 88.57 & 95.51 & 91.91
& 5.60 & 8.36 \\

DistilBERT
& 99.18 & 99.82 & 98.16 & 98.99
& 95.50 & 91.64 & 97.98 & 94.70
& 4.29 & 5.22
& 96.84 & 99.38 & 95.97 & 97.64
& 92.19 & 93.59 & 95.07 & 94.32
& 3.32 & 5.59 \\

BERT-BiLSTM
& 98.38 & 97.84 & 98.23 & 98.04
& 94.41 & 89.38 & 98.02 & 93.50
& 4.54 & 6.57
& 97.67 & 99.56 & 97.02 & 98.27
& 93.11 & 93.16 & 97.03 & 95.05
& 3.22 & 4.78 \\

TransURL
& 99.15 & 99.88 & 98.05 & 98.95
& 94.09 & 87.55 & 99.79 & 93.27
& 5.68 & 6.83
& 98.51 & 99.62 & 98.20 & 98.91
& 95.03 & 95.82 & 96.94 & 96.38
& 2.53 & 3.33 \\

PMANET
& 99.22 & 99.78 & 98.31 & 99.04
& 94.98 & 92.70 & 95.28 & 93.97
& 5.07 & 6.03
& 98.33 & 99.48 & 98.05 & 98.76
& 94.21 & 94.70 & 96.95 & 95.81
& 2.95 & 3.95 \\

\midrule

WADBERT (ours)
& \textbf{99.71} & \textbf{99.89}
& \textbf{99.39} & \textbf{99.64}
& \textbf{99.71} & \textbf{99.89}
& \textbf{99.39} & \textbf{99.64}
& \textbf{0.00} & --
& \textbf{99.44} & \textbf{99.83}
& \textbf{99.36} & \textbf{99.59}
& \textbf{99.44} & \textbf{99.83}
& \textbf{99.36} & \textbf{99.59}
& \textbf{0.00} & -- \\

\bottomrule
\end{tabular}%
}
\end{table*}

\noindent \textbf{Datasets.} We evaluate WADBERT on two public datasets, CSIC2010 and SR-BH2020.
After preprocessing and deduplication, CSIC2010 contains 36,000 benign and 25,065 malicious HTTP requests, while SR-BH2020 contains 161,334 benign and 345,942 malicious HTTP requests. The deduplicated samples are then randomly split into training and test sets at a ratio of 7:3.

\noindent \textbf{Baselines.} We compare WADBERT with six representative baselines: two traditional deep learning models and four using Transformer encoders. CNN-BiLSTM~\cite{zhou2025webguard} extracts convolutional and bidirectional recurrent features from character embeddings for classification, whereas EDL~\cite{Luo_2021} processes TF-IDF and CBOW features through MRN, LSTM, and CNN branches and fuses their outputs with an MLP classifier. DistilBERT~\cite{bokolo2023detection} uses the \texttt{[CLS]} representation of a DistilBERT encoder for classification through a fully connected layer. BERT-BiLSTM~\cite{ramos2022logbert} combines BERT contextual representations with a BiLSTM and an MLP classifier. TransURL~\cite{liu2024transurl} combines multi-layer CharBERT~\cite{ma2020charbert} features with dilated convolutions and spatial pyramid attention, followed by average pooling and linear classification. PMANET~\cite{liu2025pmanet} refines CharBERT features using CBAM-based~\cite{woo2018cbam} channel attention and spatial pyramid pooling before classification through a fully connected network. In our evaluation, all baselines receive concatenated URL--payload sequences and are assessed on the same test requests using the same metrics. This setup evaluates the benefit of WADBERT's explicit modeling of payload parameters as an unordered collection.

\noindent \textbf{Evaluation Metrics.} We use four metrics to assess request-level detection, treating malicious requests as positive instances. \textbf{(1)} Accuracy (Acc.): the fraction of requests whose labels are correctly predicted. \textbf{(2)} Precision (Pre.): the fraction of requests predicted as malicious that are actually malicious. \textbf{(3)} Recall (Rec.): the fraction of malicious requests correctly detected. \textbf{(4)} F1-score (F1): the harmonic mean of precision and recall.

\noindent \textbf{Implementation Details.}
\label{subsubsec:implementation}
We implement WADBERT and all baselines in PyTorch 2.0 with Python 3.9 on an NVIDIA A100 GPU with 64 GB RAM. WADBERT is trained for 10 epochs using the AdamW optimizer with a batch size of 8. The initial learning rate is set to 1e-5, with a 1\% linear warmup followed by a linear decay schedule. The URLBERT and SecBERT encoders adopt default BERT-base~\cite{devlin2019}
configurations with vocabulary sizes of 5K and 52K, respectively. The HGE module employs character-level embeddings (dimension = 768, vocab size = 416) and a single-layer Bi-GRU (hidden size = 384). The parameter-level attention uses 12 attention heads (head size = 64, intermediate size = 3072) with GELU activation.


\subsection{Main Results}
We compare WADBERT with six baselines to evaluate its detection performance and robustness against parameter permutations. Table~\ref{tab:overall-performance} presents the results on CSIC2010 and SR-BH2020. On the original test sets, traditional deep learning methods generally perform worse than Transformer-based methods. This suggests that stronger global contextual modeling is beneficial for web attack detection. Among all methods, WADBERT achieves the best performance on both datasets. 
On CSIC2010, it obtains an accuracy of 99.71\% and an F1-score of 99.64\%. Compared with the strongest baseline (PMANET), the improvements are 0.49\% and 0.60\%, respectively. On the more challenging SR-BH2020 dataset, WADBERT achieves 99.44\% accuracy and a 99.59\% F1-score, with an improvement of 0.93\% in accuracy and 0.68\% in F1-score over the strongest baseline (TransURL). These results demonstrate the effectiveness of WADBERT for web attack detection.

To further evaluate robustness against parameter permutations, we construct reversed test sets by completely reversing the order of payload parameters in each HTTP request. This represents an extreme order perturbation, as it reverses the relative order of every pair of parameters. We then evaluate each model on the reversed test sets using the same checkpoint as for the original test sets. As shown in Table~\ref{tab:overall-performance}, all six baselines suffer performance degradation, with F1-score decreases of 4.29--11.40 percentage points on CSIC2010 and 2.53--8.22 percentage points on SR-BH2020.  One possible explanation is that these sequential models may treat changes in parameter order as semantic changes. This can lead to different predictions for requests with the same attack logic. 
In contrast, WADBERT maintains the same detection performance before and after parameter reversal on both datasets, with no decrease in F1-score. This stability is consistent with its position-free parameter-level fusion and mean pooling, which produce a permutation-invariant payload feature (see Appendix~\ref{app:permutation_invariance} for the proof). Moreover, WADBERT outperforms the strongest baselines on the reversed test sets by 5.22\% and 3.33\% in relative F1-score on CSIC2010 and SR-BH2020, respectively. These results demonstrate WADBERT's robustness against parameter permutations and highlight the importance of permutation-invariant payload modeling for web attack detection. 
\begin{table}[t]
\centering
\caption{Detection-efficiency trade-off. F1 is reported in percentage, latency (Lat.) in milliseconds per sample, and throughput (Thr.) in samples per second.}
\begin{tabular*}{\columnwidth}{@{\extracolsep{\fill}}lcccc@{}}
\toprule
Method & F1 & GFLOPs & Lat. & Thr. \\
\midrule
\multicolumn{5}{l}{\textit{CSIC2010}} \\
CNN-BiLSTM & 96.00 & 0.23 & 0.392 & 2550.13 \\
EDL & 98.93 & 0.41 & 0.583 & 1715.63 \\
DistilBERT & 98.99 & 4.52 & 1.278 & 782.35 \\
BERT-BiLSTM & 98.04 & 9.25 & 2.424 & 412.52 \\
TransURL & 98.95 & 14.27 & 5.669 & 176.39 \\
PMANET & 99.04 & 27.59 & 7.396 & 135.21 \\
WADBERT (ours) & \textbf{99.64} & 10.86 & 2.695 & 371.02 \\
\midrule
\multicolumn{5}{l}{\textit{SR-BH2020}} \\
CNN-BiLSTM & 97.24 & 0.59 & 0.988 & 1012.54 \\
EDL & 97.51 & 1.36 & 1.446 & 691.45 \\
DistilBERT & 97.64 & 12.24 & 2.849 & 350.98 \\
BERT-BiLSTM & 98.27 & 25.08 & 5.763 & 173.52 \\
TransURL & 98.91 & 78.82 & 10.508 & 95.16 \\
PMANET & 98.76 & 110.44 & 17.781 & 56.24 \\
WADBERT (ours) & \textbf{99.59} & 63.86 & 8.073 & 123.86 \\
\bottomrule
\end{tabular*}
\label{tab:efficiency-summary}
\end{table}

\subsection{Efficiency Evaluation}
Table~\ref{tab:efficiency-summary} compares the computational cost of different methods on CSIC2010 and SR-BH2020. We report F1, GFLOPs, inference latency, and throughput. Lightweight models have lower computational costs but weaker detection performance. Transformer-based methods generally achieve better performance at a higher cost. On the CSIC2010 dataset, WADBERT is more efficient than TransURL and PMANET. It requires 10.86 GFLOPs per sample and achieves an inference latency of 2.695 ms, yielding a throughput of 371.02 samples/s. The efficiency gain mainly comes from HGE. TransURL and PMANET process an additional character-level stream through the Transformer encoder (CharBERT-based architecture \cite{ma2020charbert}). In contrast, HGE incorporates character-level information only at the embedding stage. This avoids processing a separate character stream through the full encoder. Similar results are observed on SR-BH2020. WADBERT achieves lower GFLOPs and inference latency than TransURL and PMANET while maintaining better detection performance. Its computational cost is higher on SR-BH2020 than on CSIC2010 due to the longer input sequences. Overall, WADBERT achieves a favorable trade-off between detection performance and computational efficiency on both datasets.

\subsection{Ablation Studies}
We conduct ablation studies to evaluate three key designs in WADBERT: the embedding method, the dual-channel architecture, and permutation-invariant parameter modeling.  
\begin{figure}[!t]
	\centering
    \includegraphics[width=\linewidth]{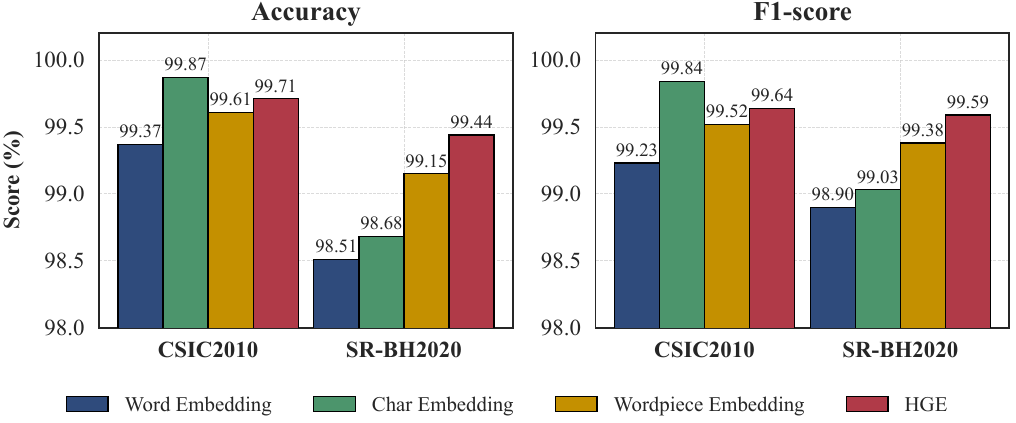}
    \vspace{-0.20in}
	\caption{Performance comparison of different embedding methods.}
	\label{fig.embedding-method-comparison}
    \vspace{-0.10in}
    \Description{}
\end{figure}
\begin{table}[t]
\centering
\caption{Computational efficiency of different embedding methods. Training time (TT) is reported in seconds per epoch, latency (Lat.) in milliseconds per sample, and throughput (Thr.) in samples per second.}
\begin{tabular*}{\columnwidth}{@{\extracolsep{\fill}}llcccc@{}}
\toprule
Dataset & Method & TT (s) & GFLOPs & Lat. & Thr. \\
\midrule
\multirow{4}{*}{CSIC2010}
& Word & \textbf{180.34} & \textbf{8.12}
& \textbf{1.608} & \textbf{621.91} \\
& WordPiece & 182.89 & 8.15 & 1.909 & 523.88 \\
& Char & 360.56 & 20.76 & 3.753 & 266.45 \\
& HGE & 244.72 & 10.86 & 2.695 & 371.02 \\
\midrule
\multirow{4}{*}{SR-BH2020}
& Word & \textbf{5672.12} & \textbf{43.86}
& \textbf{6.259} & \textbf{159.77} \\
& WordPiece & 7665.87 & 50.92 & 6.779 & 147.51 \\
& Char & 16527.30 & 113.14 & 16.828 & 59.42 \\
& HGE & 9135.51 & 63.86 & 8.073 & 123.86 \\
\bottomrule
\end{tabular*}
\label{tab:embedding-efficiency}
\end{table}

\noindent \textbf{Impact of Embedding Methods.}
We compare four embedding strategies in Figure~\ref{fig.embedding-method-comparison} and Table~\ref{tab:embedding-efficiency}. On the CSIC2010 dataset, Char embedding achieves the highest F1-score of 99.84\%, followed by HGE at 99.64\%. However, HGE requires 32.1\% less training time and 47.7\% fewer GFLOPs than Char embedding. Moreover, HGE achieves improvements of 0.10\% in accuracy and 0.12\% in F1-score over WordPiece embedding. In contrast, word embedding performs the worst, reflecting its limited ability to represent the symbolic and irregular patterns appearing in HTTP requests. 
On SR-BH2020, HGE achieves the best accuracy of 99.44\% and F1-score of 99.59\%. These scores surpass WordPiece embedding by 0.29\% and 0.21\%, respectively. Notably, Char embedding exhibits a performance decline. This suggests that pure character-level representations fall short of capturing richer semantic dependencies and contextual patterns. Moreover, HGE requires 44.7\% less training time and 43.6\% fewer GFLOPs than Char embedding. The reason is that Char embedding represents individual characters as tokens, resulting in longer input sequences. In contrast, HGE uses a lightweight GRU to incorporate character features efficiently, maintaining fine-grained details without extending token sequences. Overall, these results demonstrate that HGE achieves a favorable trade-off between detection performance and computational efficiency.

\noindent \textbf{Impact of Model Components.}
\begin{table}[t]
\centering
\caption{Ablation study of the dual-channel architecture (\%).}
\label{tab:module-ablation}
\begin{tabular*}{\columnwidth}{@{\extracolsep{\fill}}llcccc@{}}
\toprule
Dataset & Model Variant & Acc. & Pre. & Rec. & F1 \\
\midrule
\multirow{3}{*}{CSIC2010}
& w/o SecBERT & 73.63 & 61.06 & 98.54 & 75.40 \\
& w/o URLBERT & 90.30 & 99.78 & 76.60 & 86.67 \\
& WADBERT & \textbf{99.71} & \textbf{99.89}
          & \textbf{99.39} & \textbf{99.64} \\
\midrule
\multirow{3}{*}{SR-BH2020}
& w/o SecBERT & 94.83 & 97.81 & 94.54 & 96.15 \\
& w/o URLBERT & 85.01 & 85.33 & 94.21 & 89.55 \\
& WADBERT & \textbf{99.44} & \textbf{99.83}
          & \textbf{99.36} & \textbf{99.59} \\
\bottomrule
\end{tabular*}
\end{table}
We construct two single-channel variants to evaluate the dual-channel architecture. WADBERT w/o SecBERT uses only URL information, and WADBERT w/o URLBERT uses only payload information. Table~\ref{tab:module-ablation} compares them with WADBERT, which uses both modalities. On CSIC2010, the payload-only model performs better than the URL-only model, with F1-scores of 86.67\% and 75.40\%, respectively. The reason is that payload parameters contain rich attack features, such as \texttt{alert('XSS')} and \texttt{OR' 1=1}. However, the payload-only model cannot detect attacks occurring in URL paths (e.g., path traversal attacks). The opposite trend is observed on SR-BH2020, where the URL-only model performs better. This is because SR-BH2020 contains more RESTful API requests (e.g., \path{/blog/{id}/uploads/{file}}), where attackers can inject malicious payloads directly into URL paths (e.g., \path{/blog/alert('XSS');/uploads/test.jpg}). These results confirm that URL and payload information are complementary. By combining both channels, WADBERT captures abnormal URL paths and malicious payload parameters.

\noindent \textbf{Impact of Permutation-Invariant Parameter Modeling.} We evaluate two key designs for permutation-invariant parameter modeling: the parameter-level fusion strategy and the exclusion of positional encoding. To assess the former, we introduce FlatPayload, which concatenates all parameters sequentially into a single text input for SecBERT encoding and removes the parameter-level attention module. To examine the latter, we introduce WADBERT with positional encoding (w/ PE), a variant that adds positional encoding to the parameter-level attention module and keeps all other components unchanged. 
Table~\ref{tab:payload_pe_ablation} shows that WADBERT outperforms FlatPayload on both datasets, improving the F1-score by 8.30\% and 3.98\% on CSIC2010 and SR-BH2020, respectively. FlatPayload also exhibits substantially lower precision, indicating more false positives. One possible explanation is that sequential models (e.g., FlatPayload) may treat parameter permutations as semantic changes, leading to misclassifications even when the attack logic remains unchanged. In contrast, WADBERT captures combinatorial relationships among parameters through parameter-level fusion without relying on their order. 
Furthermore, WADBERT achieves F1-score improvements of 3.02\% and 0.61\% over WADBERT w/ PE on CSIC2010 and SR-BH2020, respectively. This may be because positional encoding introduces unnecessary sensitivity to parameter order. Overall, the findings support the effectiveness of WADBERT's permutation-invariant parameter modeling design.
\begin{table}[t]
\centering
\caption{Ablation study of parameter-level fusion and positional encoding for permutation-invariant parameter modeling (\%).}
\begin{tabular*}{\columnwidth}{@{\extracolsep{\fill}}llcccc@{}}
\toprule
Dataset & Method & Acc. & Pre. & Rec. & F1 \\
\midrule
\multirow{3}{*}{CSIC2010}
& FlatPayload & 92.17 & 84.52 & 99.35 & 91.34 \\
& WADBERT w/ PE & 97.15 & 94.02 & 99.36 & 96.62 \\
& WADBERT & \textbf{99.71} & \textbf{99.89}
& \textbf{99.39} & \textbf{99.64} \\
\midrule
\multirow{3}{*}{SR-BH2020}
& FlatPayload & 93.79 & 92.39 & 99.05 & 95.61 \\
& WADBERT w/ PE & 98.61 & 98.78 & 99.19 & 98.98 \\
& WADBERT & \textbf{99.44} & \textbf{99.83}
& \textbf{99.36} & \textbf{99.59} \\
\bottomrule
\end{tabular*}
\label{tab:payload_pe_ablation}
\end{table}

\subsection{Malicious Parameter Localization.}

\noindent \textbf{Dataset and Evaluation Setup.}
We evaluate malicious parameter localization using WebSpotter's manually annotated CSIC2010 dataset~\cite{li2026achieving} under two settings. First, we match the attack requests in our test set against all annotated requests released by WebSpotter. This yields 5,749 matched attack requests (33,927 parameters) for evaluating WADBERT's malicious parameter localization ability. Second, to compare WADBERT with existing localization methods, we restrict the matching to WebSpotter's official test set. This avoids potential data leakage caused by different training and test splits. Under this setting, we obtain 1,576 matched attack requests (8,654 parameters). We then evaluate WADBERT on this subset and use the published results of WebSpotter and its baselines as reference points, since the evaluation sets are not identical. For both settings, we formulate malicious parameter localization as a binary classification task, where each parameter is labeled as malicious or benign. Following WebSpotter's evaluation method, we compute precision, recall, F1-score, and Jaccard Index separately for each request and then average the results across all evaluated requests.

\noindent \textbf{Attention-Based Localization.}
We use attention weights from request classification to identify potentially malicious parameters. The received-attention score $s_k$ is defined in Eq.~\eqref{eq:parameter-importance}: it averages the weights $A_{\ell k}^{(h)}$ over all $H$ attention heads and $n$ query parameters. Here, $A_{\ell k}^{(h)}$ is the $(\ell,k)$ entry of the attention matrix $\mathbf{A}^{(h)}$, representing the attention from parameter $\ell$ to parameter $k$. Since the scores satisfy $\sum_{k=1}^{n}s_k=1$, we use $1/n$ as the request-adaptive threshold and identify parameters with $s_k\geq1/n$ as potentially malicious. This strategy allows WADBERT to localize multiple malicious parameters without training an additional localization model.
\begin{table}[t]
\centering
\caption{Reference comparison of malicious-parameter localization performance (\%). Baseline results are reported by WebSpotter on its full test set, while WADBERT is evaluated on 1,576 matched attack requests. }
\begin{tabular*}{\columnwidth}{@{\extracolsep{\fill}}lcccc@{}}
\toprule
Method & Pre. & Rec. & F1 & Jaccard Index \\
\midrule
LTD \cite{liu2019locate} & 42.00 & 77.60 & 52.00 & 42.00 \\
SSD \cite{liu2016ssd} & 26.60 & \textbf{99.80} & 40.10 & 27.10 \\
REG \cite{noothout2020deep} & 10.20 & 23.50 & 12.90 & 10.20 \\
CLS \cite{lampert2008beyond} & 86.70 & 96.90 & 90.00 & 86.70 \\
WebSpotter \cite{li2026achieving} & 98.00 & 98.70 & 98.20 & 98.00 \\
WADBERT (ours) & \textbf{98.48} & 98.86 & \textbf{98.60} & \textbf{98.48} \\
\bottomrule
\end{tabular*}

\label{tab:parameter-localization}
\end{table}

\noindent \textbf{Localization Results and Analysis.} In the first evaluation setting, we evaluate WADBERT on 5,749 attack requests from our own test set. WADBERT achieves a localization precision of 98.05\%, recall of 98.52\%, F1-score of 98.21\%, and Jaccard Index of 98.05\% on 5,749 matched attack requests. The high precision and recall indicate that WADBERT can accurately identify malicious parameters while minimizing false positives and missed detections. A possible explanation is that WADBERT independently extracts semantic features from each parameter and uses parameter-level attention to assign higher attention weights to suspicious parameters. 

In the second setting, we compare WADBERT with existing localization methods, as shown in Table~\ref{tab:parameter-localization}. On the 1,576 attack requests matched with WebSpotter's test set, WADBERT achieves a precision of 98.48\%, recall of 98.86\%, F1-score of 98.60\%, and Jaccard Index of 98.48\%. For reference, WebSpotter reports an F1-score of 98.20\% on its full test set. Since these evaluation sets are not identical, the published baseline results are provided as reference points rather than directly comparable measurements. Together with the 98.21\% F1-score obtained on 5,749 matched requests from our own test set, these results provide complementary evidence of WADBERT's malicious-parameter localization capability. Notably, WADBERT achieves malicious-parameter localization without requiring additional parameter-level annotations for training. Instead, it directly leverages the attention weights learned during classification to highlight potentially malicious parameters. To further examine the localization behavior, we present three representative examples involving XSS, SQLi, and composite attacks. Supplementary experiments are provided in Appendix~\ref{app:supplementary-experiments}.

\section{Conclusion}
\label{sec:conclusion}
We propose WADBERT, a structure-aware web attack detection model that achieves high accuracy while improving the transparency of detection results. It employs HGE to combine subword semantics with character-level information, improving the representation of irregular and obfuscated HTTP requests. To avoid unnecessary sensitivity to payload parameter order, WADBERT employs a permutation-invariant modeling strategy. The learned attention weights further provide parameter-level diagnostic cues. Experiments on CSIC2010 and SR-BH2020 show that WADBERT achieves accuracies of 99.71\% and 99.44\%, respectively. Ablation studies further confirm the effectiveness of HGE, the dual-channel architecture, and parameter-level fusion. Future work will evaluate WADBERT on larger real-world datasets and investigate adversarial training to improve its generalization and robustness.


\bibliographystyle{sections/ACM-Reference-Format}
\bibliography{sections/reference}
\appendix
\section{Permutation Invariance of Payload Feature Extraction}
\label{app:permutation_invariance}
\begin{figure*}[t]
    \centering
    \includegraphics[width=\textwidth]{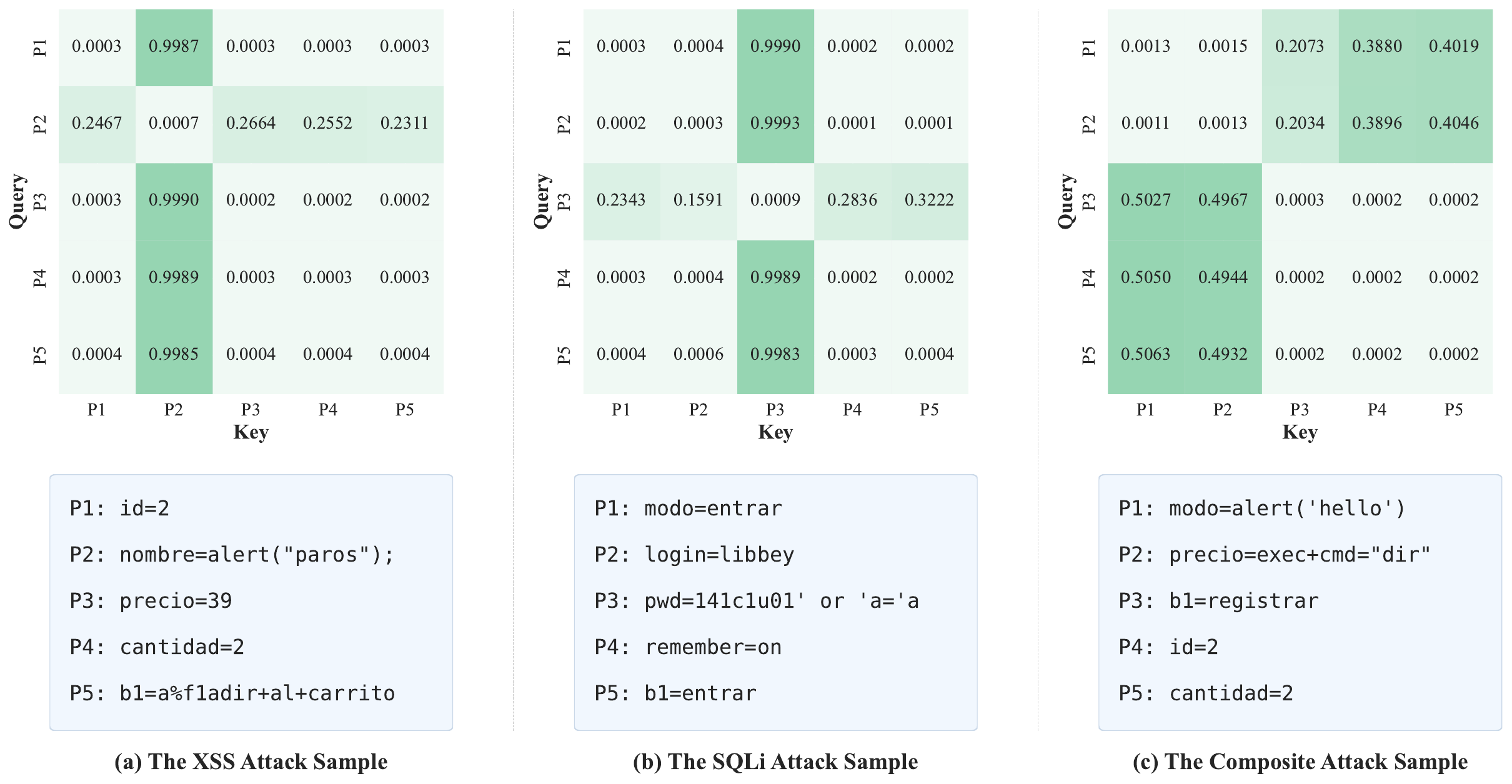}
    \caption{Parameter-level attention distributions for representative attack samples.}
    \label{fig:parameter-attention}
    \Description{Attention heatmaps for XSS, SQL injection, and composite attack examples.}
\end{figure*}
We prove that the aggregation map $\rho$ defined in Section~\ref{sec:payload-feature-extraction} is invariant under permutations of parameter order. All statements concern deterministic inference on a fixed collection of $n\geq1$ retained parameters.

\noindent\textbf{Proposition 1.}
Let $\mathbf{P}\in\mathbb{R}^{n\times d}$ be the parameter-level feature matrix, and let
$\mathbf{R}\in\{0,1\}^{n\times n}$ be a permutation matrix. Multi-head self-attention without parameter-level positional encoding satisfies
\begin{equation}
\operatorname{MHA}(\mathbf{R}\mathbf{P})
=\mathbf{R}\operatorname{MHA}(\mathbf{P}).
\label{eq:attention_equivariance}
\end{equation}
Consequently, mean pooling produces a permutation-invariant payload feature:
\begin{equation}
\rho(\mathbf{R}\mathbf{P})=\rho(\mathbf{P}).
\end{equation}

\noindent\textbf{Proof.}
Consider attention head $h$. Its query, key, and value matrices are
$\mathbf{Q}^{(h)}=\mathbf{P}\mathbf{W}_{\mathrm{Q}}^{(h)}$,
$\mathbf{K}^{(h)}=\mathbf{P}\mathbf{W}_{\mathrm{K}}^{(h)}$, and
$\mathbf{V}^{(h)}=\mathbf{P}\mathbf{W}_{\mathrm{V}}^{(h)}$,
where $\mathbf{W}_{\mathrm{Q}}^{(h)},\mathbf{W}_{\mathrm{K}}^{(h)}\in\mathbb{R}^{d\times d_{\mathrm{k}}}$ and
$\mathbf{W}_{\mathrm{V}}^{(h)}\in\mathbb{R}^{d\times d_{\mathrm{v}}}$ are learnable projections. Here, $d_{\mathrm{k}}$ and $d_{\mathrm{v}}$ are the key and value dimensions of each head. Define the score and attention matrices by
\begin{equation}
\begin{aligned}
\mathbf{S}^{(h)}
&\coloneqq\frac{\mathbf{Q}^{(h)}(\mathbf{K}^{(h)})^\top}
{\sqrt{d_{\mathrm{k}}}},\\
\mathbf{A}^{(h)}
&\coloneqq\operatorname{softmax}_{\mathrm{row}}
(\mathbf{S}^{(h)}),
\end{aligned}
\end{equation}
where $\operatorname{softmax}_{\mathrm{row}}$ normalizes each row. For brevity, we omit the head superscript in the remaining single-head derivation.

After permuting the parameters, the projections become
$\mathbf{Q}'=\mathbf{R}\mathbf{Q}$,
$\mathbf{K}'=\mathbf{R}\mathbf{K}$, and
$\mathbf{V}'=\mathbf{R}\mathbf{V}$. Therefore,
\begin{equation}
\begin{aligned}
\mathbf{S}'
&=\frac{(\mathbf{R}\mathbf{Q})(\mathbf{R}\mathbf{K})^\top}
{\sqrt{d_{\mathrm{k}}}}\\
&=\mathbf{R}\mathbf{S}\mathbf{R}^\top.
\end{aligned}
\end{equation}

Right multiplication by $\mathbf{R}^\top$ permutes the entries within each row, preserving the softmax normalization term. Left multiplication by $\mathbf{R}$ permutes the rows. Hence,
\begin{equation}
\begin{aligned}
\mathbf{A}'
&=\operatorname{softmax}_{\mathrm{row}}
(\mathbf{R}\mathbf{S}\mathbf{R}^\top)\\
&=\mathbf{R}\mathbf{A}\mathbf{R}^\top.
\end{aligned}
\label{eq:softmax_permutation}
\end{equation}
A single attention head outputs
$\operatorname{Att}(\mathbf{P})=\mathbf{A}\mathbf{V}$. Using
$\mathbf{R}^\top\mathbf{R}=\mathbf{I}_n$, where $\mathbf{I}_n$ is the $n\times n$ identity matrix, we obtain
\begin{equation}
\begin{aligned}
\operatorname{Att}(\mathbf{R}\mathbf{P})
&=\mathbf{A}'\mathbf{V}'\\
&=\mathbf{R}\mathbf{A}\mathbf{R}^\top
\mathbf{R}\mathbf{V}\\
&=\mathbf{R}\operatorname{Att}(\mathbf{P}).
\end{aligned}
\end{equation}
Each head is thus permutation-equivariant. Head concatenation and the output projection act along the feature dimension and commute with row permutations, establishing Eq.~\eqref{eq:attention_equivariance}.

Finally, applying the mean pooling definition in Eq.~\eqref{eq:mean-pooling} yields
\begin{equation}
\begin{aligned}
\rho(\mathbf{R}\mathbf{P})
&=\frac{1}{n}
\operatorname{MHA}(\mathbf{R}\mathbf{P})^\top\mathbf{1}_n\\
&=\frac{1}{n}
\operatorname{MHA}(\mathbf{P})^\top
\mathbf{R}^\top\mathbf{1}_n\\
&=\frac{1}{n}
\operatorname{MHA}(\mathbf{P})^\top\mathbf{1}_n\\
&=\rho(\mathbf{P}),
\end{aligned}
\end{equation}
because $\mathbf{R}^\top\mathbf{1}_n=\mathbf{1}_n$. Therefore, the payload feature
$\mathbf{f}_{\mathrm{payload}}=\rho(\mathbf{P})$
is permutation-invariant. This completes the proof.

\begin{table*}[t]
\centering
\caption{Complete adversarial attack results. ASR, post-attack accuracy (Acc.), and perturbation rate (Per.) are percentages. Que. is the average number of victim-model queries.}
\label{tab:adversarial-full}
\begin{tabular*}{\textwidth}{@{\extracolsep{\fill}}lrrrrrrrrrr@{}}
\toprule
& \multicolumn{5}{c}{CSIC2010} & \multicolumn{5}{c}{SR-BH2020} \\
\cmidrule(lr){2-6}\cmidrule(lr){7-11}
Attack & ASR & Acc. & Que. & Per. & USE & ASR & Acc. & Que. & Per. & USE \\
\midrule
TextFooler  & 15.83 & 83.92 & 52.31  & 7.05  & 0.9209 & 18.60 & 80.81 & 95.82  & 2.00  & 0.9630 \\
BERT-Attack & 16.62 & 83.13 & 107.81 & 8.30  & 0.9215 & 26.50 & 72.96 & 178.44 & 2.72  & 0.9678 \\
BAE-R       & 20.73 & 79.03 & 19.32  & 8.33  & 0.8652 & 22.86 & 76.57 & 138.70 & 2.53  & 0.9648 \\
CLARE       & 24.47 & 75.31 & 21.39  & 36.84 & 0.8358 & 31.30 & 68.19 & 150.00 & 28.45 & 0.9639 \\
A2T-MLM     & 35.10 & 64.71 & 6.93   & 10.81 & 0.8636 & 36.30 & 63.23 & 56.72  & 4.36  & 0.9600 \\
\bottomrule
\end{tabular*}
\begin{minipage}[t]{\textwidth}
\smallskip
\small
\begin{enumerate}[label=\arabic*., itemsep=0pt, leftmargin=*, labelindent=0pt]
    \item ASR denotes the attack success rate; Acc denotes the classification accuracy of the victim-model under attack.
    \item Queries denotes the average number of victim-model queries per adversarial example; Per. denotes the average perturbation rate.
    \item USE score denotes the average semantic similarity between the original and adversarial examples, where the similarity for each pair is measured by USE.\\
\end{enumerate}
\end{minipage}
\end{table*}

\begin{table}[t]
\caption{Cross-domain adaptation performance (\%).}
\label{tab:cross-domain-adaptation}
\centering
\begin{tabular*}{\columnwidth}{@{\extracolsep{\fill}}lcc@{}}
\toprule
Metric &
\shortstack{CSIC2010 $\rightarrow$ SR-BH2020} &
\shortstack{SR-BH2020 $\rightarrow$ CSIC2010} \\
\midrule
Accuracy  & 92.36 & 96.03 \\
Precision & 95.27 & 97.02 \\
Recall    & 93.46 & 93.19 \\
F1-score  & 94.36 & 95.07 \\
\bottomrule
\end{tabular*}
\end{table}
\section{Supplementary Experiments}
\label{app:supplementary-experiments}
\subsection{Attention Visualization for Malicious Parameter Localization}
\label{app:attention-visualization}

Figure~\ref{fig:parameter-attention} provides three qualitative examples that complement the localization metrics reported in Section~\ref{sec:evaluation}. In the XSS example, parameter $p_2$, which contains \texttt{alert}, receives an importance score of $0.7991$, compared with $0.0496$, $0.0535$, $0.0513$, and $0.0465$ for the other four parameters. In the SQLi example, $p_3$, which contains the SQL fragment \texttt{or 'a='a'}, receives the highest attention. In the composite example, both $p_1$ (XSS) and $p_2$ (command injection) receive high attention weights. These cases illustrate how the learned attention scores can help identify parameters for further inspection; they do not, by themselves, establish causal explanations for the model's predictions.

\subsection{Adversarial Robustness Evaluation.}
\label{app:adversarial}
We evaluate four black-box attacks, including TextFooler~\cite{jin2020textfooler}, BERT-Attack~\cite{li2020BERT-ATTACK}, BAE-R~\cite{garg2020BAE}, and CLARE~\cite{li2021CLARE}. We also evaluate the white-box attack A2T-MLM~\cite{yoo2021A2T}. TextFooler uses embedding-neighbor substitutions; BERT-Attack, BAE-R, CLARE, and A2T-MLM use contextual substitutions, with CLARE additionally supporting insertion and merging operations. A2T-MLM uses gradient-based word-importance ranking and therefore assumes stronger access to the victim model than the black-box methods. Table~\ref{tab:adversarial-full} reports the adversarial attack results.

\noindent\textbf{Attack effectiveness.}
A2T-MLM achieves the highest ASR on both datasets. A possible explanation is that its gradient-based word-importance ranking prioritizes words that strongly affect model predictions. In contrast, TextFooler achieves the lowest ASR because its nearest-neighbor search produces less effective substitutions. Although these attacks degrade the performance of WADBERT, it still achieves post-attack accuracies ranging from 63.23\% to 83.92\%. Overall, WADBERT retains reasonable robustness against the black-box attacks. However, its robustness remains limited against the gradient-based white-box attack A2T-MLM.

\noindent\textbf{Perturbation Quality.}
TextFooler yields perturbation rates of $7.05\%$ on CSIC2010 and $2.00\%$ on SR-BH2020. BERT-Attack obtains USE similarity scores of $0.9215$ and $0.9678$, respectively. In comparison, the stronger attacks tend to use larger perturbations and yield lower similarity scores. These measurements characterize the evaluated text perturbations but do not, on their own, guarantee preservation of valid HTTP syntax or attack semantics.

\noindent\textbf{Attack Efficiency.}
A2T-MLM uses an average of $6.93$ and $56.72$ victim-model queries on CSIC2010 and SR-BH2020, respectively. Because A2T-MLM also uses gradient information, its query count should not be interpreted as directly comparable to the query budgets of black-box methods. Among the black-box methods, TextFooler requires more queries than BAE-R and CLARE on CSIC2010, whereas the reverse pattern is observed on SR-BH2020. This difference suggests that search cost depends on the dataset and on the candidate-substitution strategy.

\subsection{Cross-Domain Adaptation Evaluation.}

We evaluate our model under two cross-domain adaptation settings. In each setting, WADBERT is first trained on the source dataset and then fine-tuned with limited labeled data from the target dataset (2,000 samples). It is finally evaluated on the target test set. As reported in Table~\ref{tab:cross-domain-adaptation}, WADBERT achieves F1-scores of 94.36\% for CSIC2010 $\rightarrow$ SR-BH2020 and 95.07\% for SR-BH2020 $\rightarrow$ CSIC2010. Compared with the in-domain results in Table~\ref{tab:overall-performance}, performance drops under cross-domain evaluation, since the two datasets differ in attack patterns and request structures. In SR-BH2020, attack-bearing content is typically embedded in URL paths, whereas in CSIC2010, it is typically found in query strings or request bodies. Overall, these results show that WADBERT can adapt to a new domain with limited labeled target-domain data.
\subsection{Hyperparameter Sensitivity}
\label{app:hyperparameters}

\begin{figure*}[t]
    \centering
    \includegraphics[width=0.9\textwidth]{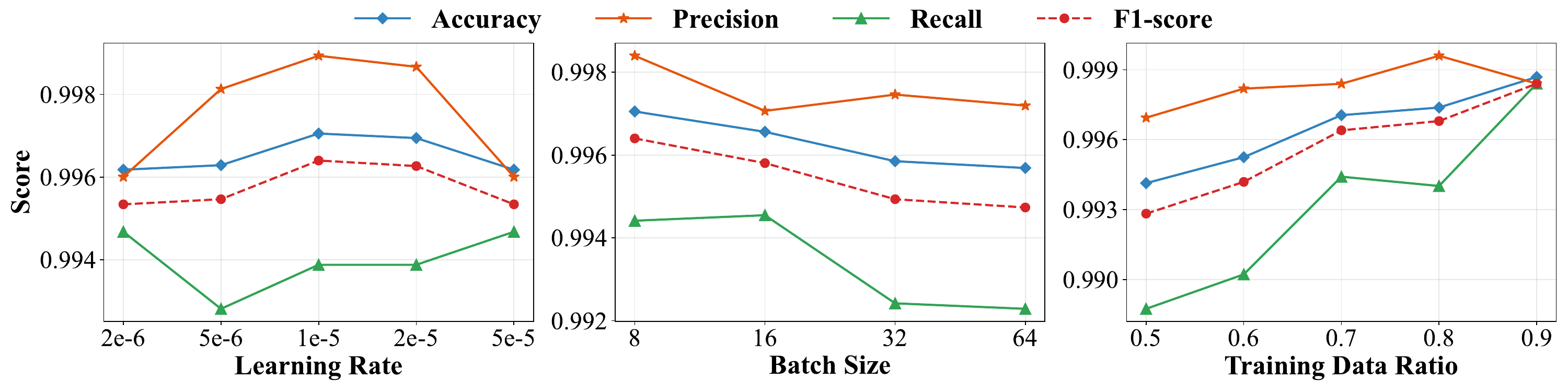}
    \includegraphics[width=0.9\textwidth]{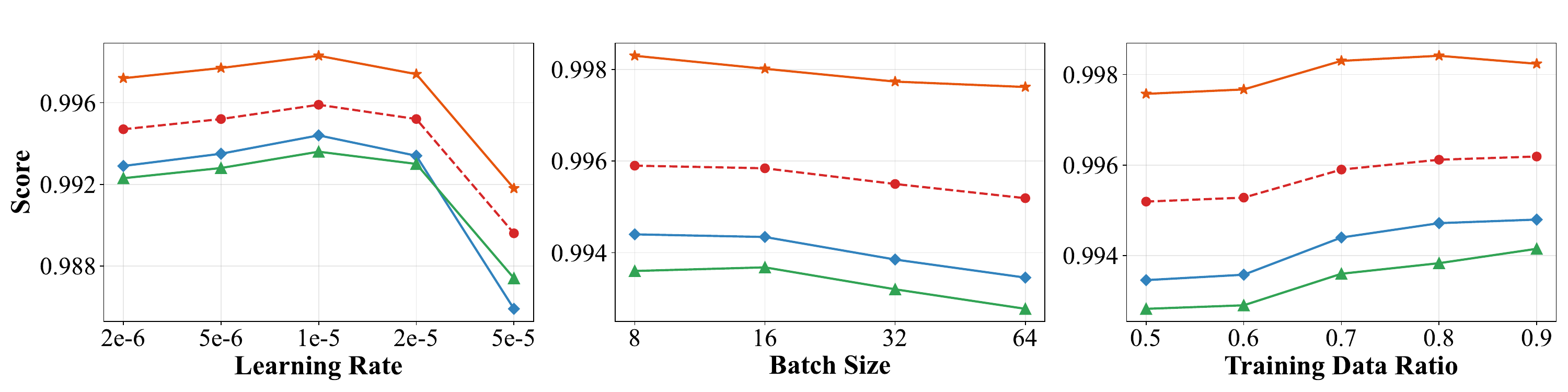}
    \caption{Performance of WADBERT under different hyperparameter settings on CSIC2010 (top) and SR-BH2020 (bottom).}
    \label{fig:hyperparameter}
    \Description{Hyperparameter sensitivity curves for the learning rate, batch size, and training data ratio on the two datasets.}
\end{figure*}

Figure~\ref{fig:hyperparameter} provides the detailed curves underlying the trends summarized in the main text. We discuss the differences among the three hyperparameters below.

\noindent\textbf{Learning Rate.}
On SR-BH2020, reducing the learning rate from $5\times10^{-5}$ to $10^{-5}$ improves performance, whereas further reductions to $5\times10^{-6}$ and $2\times10^{-6}$ lead to small declines. A similar pattern occurs on CSIC2010. Values between $5\times10^{-6}$ and $2\times10^{-5}$ yield relatively stable results. The weaker performance at $5\times10^{-5}$ is consistent with overly large parameter updates, while smaller learning rates may slow convergence within the fixed training budget.

\noindent\textbf{Batch Size.}
Although overall accuracy and F1-score vary only slightly across the evaluated batch sizes, recall is more sensitive than precision, particularly on CSIC2010. This indicates that changes in batch size may affect the detection of some malicious requests even when aggregate performance remains similar.

\noindent\textbf{Training Data Ratio.}
The performance gain from additional training samples is smaller on SR-BH2020 than on CSIC2010. In particular, the SR-BH2020 curves approach saturation at a training data ratio of $0.5$, whereas CSIC2010 continues to benefit more noticeably from additional samples. This difference suggests that the two datasets require different amounts of labeled data to approach their observed performance ceilings.

\end{document}